\documentclass[aps,prl,twocolumn,superscriptaddress]{revtex4-2}
\usepackage{amsmath,graphicx}
\usepackage[noabbrev]{cleveref}
\usepackage{siunitx}
\usepackage{siunitx}
\begin{document}

\title{
%Excited-states spectroscopy of a one-particle bilayer graphene quantum dot]{Excited-states spectroscopy of a one-particle bilayer graphene quantum dot from time-resolved charge detection
Spin-valley locked excited states spectroscopy in a one-particle bilayer graphene quantum dot
}


\author{Hadrien Duprez}
\author{Solenn Cances}
\author{Andraz Omahen}
\author{Michele Masseroni}
\author{Max J. Ruckriegel}
\author{Christoph Adam}
\author{Chuyao Tong}
\author{Jonas Gerber}
\author{Rebekka Garreis}
\author{Wister Huang}
\author{Lisa Gächter}
\affiliation{Solid State Physics Laboratory, ETH Zurich, CH-8093 Zurich, Switzerland}
\author{Takashi Taniguchi}
\affiliation{Research Center for Materials Nanoarchitectonics, National Institute for Materials Science,  1-1 Namiki, Tsukuba 305-0044, Japan}
\author{Kenji Watanabe}
\affiliation{Research Center for Electronic and Optical Materials, National Institute for Materials Science, 1-1 Namiki, Tsukuba 305-0044, Japan}
\author{Thomas Ihn}
\author{Klaus Ensslin}
\affiliation{Solid State Physics Laboratory, ETH Zurich, CH-8093 Zurich, Switzerland}

\begin{abstract}
Current semiconductor qubits rely either on the spin or on the charge degree of freedom to encode quantum information.
By contrast, in bilayer graphene the valley degree of freedom, stemming from the crystal lattice symmetry, is a robust quantum number that can therefore be harnessed for this purpose. 
The simplest implementation of a valley qubit would rely on two states with opposite valleys as in the case of a single-carrier bilayer graphene quantum dot immersed in a small perpendicular magnetic field ($B_\perp\lesssim 100$mT).
However, the single-carrier quantum dot excited states spectrum has not been resolved to date in the relevant magnetic field range.  
Here, we fill this gap, by measuring the parallel and perpendicular magnetic field dependence of this spectrum  with an unprecedented resolution of $4\mu$eV.
We use a time-resolved charge detection technique that gives us access to individual tunnel events.
Our results come as a direct verification of the predicted spectrum and establish a new upper-bound on inter-valley mixing, equal to our energy resolution.
Our charge detection technique opens the door to measuring the relaxation time of a valley qubit in a single-carrier bilayer graphene quantum dot. 
\end{abstract}

%\keywords{keyword1, Keyword2, Keyword3, Keyword4}



\maketitle

\section{Introduction}\label{sec1}

Quantum dots are an essential building block of electrical quantum circuits due to their diverse and versatile nature, positioning them as a primary physical platform for hosting qubits.
Currently, semiconductor qubits rely on charge or spin for encoding quantum information~\cite{chatterjee2021SemiconductorQubits,burkard2023SemiconductorSpin}.
The valley degree of freedom, which stems from the crystal lattice symmetry, stands as a possible alternative.

Spin qubits are ultimately limited by phonons and charge noise that cause relaxation or dephasing through one of the spin-orbit, exchange or hyperfine interactions~\cite{burkard2023SemiconductorSpin}. 
On the other hand, relaxation and decoherence processes for the valley qubits remain to be explored.
Bernal-stacked bilayer graphene (BLG) is an ideal material to study these effects, as the charge carriers' angular momentum is characterized by both the spin and the valley that are each robust quantum numbers.
The strong valley blockade~\cite{tong2022PauliBlockade} leading to long relaxation times of a valley singlet to a valley triplet state~\cite{garreis2024LonglivedValley} in a BLG double quantum dot showcases the valley degree of freedom as comparatively less fragile than the spin.
In a BLG single quantum dot populated with one charge carrier, the intrinsic spin--orbit interaction couples the spin- and orbital- angular momenta which consequently share the same, out-of-plane, quantization axis. 
As a result, the four-fold degeneracy is lifted and the spectrum is composed of two Kramers pairs, each consisting of states with both opposite valley and spin, noted  ($|K^-\uparrow\rangle, |K^+\downarrow\rangle$) and ($|K^-\downarrow\rangle, |K^+\uparrow\rangle$), as illustrated in figure~\ref{fig:1}f~\cite{banszerus2021SpinvalleyCoupling,kurzmann2021KondoEffect,knothe2022TunnelingTheory}. 
These pairs exemplify the spin-valley locking mechanism at play in 2D materials with spin-orbit interaction of Kane-Mele~\cite{kane2005QuantumSpin} and Ising type~\cite{lu2015EvidenceTwodimensional,saito2016SuperconductivityProtected,sierra2021VanWaals,masseroni2024SpinorbitProximity}.

In BLG quantum dots, the N-particle ground state spectrum was extensively studied~\cite{eich2018SpinValley,kurzmann2019ExcitedStates,tong2021TunableValley,garreis2021ShellFilling,moller2023UnderstandingFourfold}.
However, the existing data on the excited states spectrum of a single-particle quantum dot is more scarce despite several attempts using different measurement approaches~\cite{kurzmann2019ExcitedStates,banszerus2021PulsedgateSpectroscopy,banszerus2021SpinvalleyCoupling,garreis2022CountingStatistics}.
In particular, the excited states spectrum of a single-particle quantum dot was never measured at low perpendicular magnetic fields ($B_\perp \lesssim \SI{100}{mT}$), where the first crossing of excited states occurs\cite{knothe2020QuartetStates,knothe2022TunnelingTheory}.
In addition, only the spin- (and no valley-) excited state relaxation time was measured~\cite{banszerus2021PulsedgateSpectroscopy,banszerus2022SpinRelaxation,gachter2022SingleShotSpin}, due to insufficient energy resolution~\cite{banszerus2021PulsedgateSpectroscopy,banszerus2022SpinRelaxation} or too high tunnel rates~\cite{gachter2022SingleShotSpin,garreis2022CountingStatistics} in the appropriate magnetic field range.
Here, we overcome both of these shortcomings and experimentally establish the---so far simply assumed---single-charge carrier spectrum picture at zero and low magnetic fields.
In addition, we determine a new upper bound on the inter-valley mixing energy scale, equivalent to our energy resolution of $\SI{4}{\micro eV}$, a five-fold improvement on the previously reported upper bound~\cite{banszerus2021SpinvalleyCoupling}.
Moreover, the tunnel rates into and out of the quantum dot were tuned down to unprecedentedly low values of \SI{12.2}{Hz} at $B=\SI{0}{T}$.
We overcame the previous requirement of a finite perpendicular magnetic field to reach sub-kHz tunnel rates~\cite{gachter2022SingleShotSpin,garreis2023CountingStatistics,garreis2024LonglivedValley}, likely thanks to a relatively large displacement field ($D/\epsilon_0\approx\SI{0.9}{V/nm}$, where $\epsilon_0$ is the vacuum permittivity), which opens a bandgap larger than the disorder potential. 
Finally, the time-resolved charge detection scheme used here for the spectroscopy, in combination with our low effective electron temperature ($\SI{46}{mK}$), opens the door to measuring the relaxation time of the different excited states of the first charge carrier spectrum via single-shot readout~\cite{elzerman2004SingleshotReadout} even at low magnetic fields. 

The rate at which a charge carrier tunnels into or out of a quantum dot is given by Fermi's golden rule. This rate therefore depends on the spatial extent of the dot's orbital wavefunction, the number of available micro-states (ground and excited states), and selection rules. 
Notably, in the case of the transition from 0 to 1 carrier, there are no selection rules as the leads are neither valley- nor spin-polarized.
The simplest model assumes that all of the micro-states of a given charge state have an identical orbital wave function. 
In this case, the tunneling in (out) rate $\Gamma_\mathrm{in(out)}$ is expected to increase in steps proportional to the number of available micro-states~\cite{park2003ElectronTransport,elzerman2004SingleshotReadout, hofmann2016EquilibriumFree,hofmann2016MeasuringDegeneracy}: 
\begin{align}
    \Gamma_\mathrm{out}(\varepsilon) &= \Gamma \left[ 1-f(\varepsilon-\varepsilon_\mathrm{g}) \right] \label{eq:Gamma_ou} \\
    \Gamma_\mathrm{in}(\varepsilon) &= \Gamma \left[ f(\varepsilon-\varepsilon_\mathrm{g}) + \sum_{i} f(\varepsilon-\varepsilon_\mathrm{g}-\Delta\varepsilon_{i}) \right] \label{eq:Gamma_in}
\end{align}
where $\Gamma$ is the tunnel rate intrinsic to the barrier configuration and the orbital's spatial extent, $f$ is the Fermi-Dirac distribution, $\varepsilon_\mathrm{g}$ is the ground state energy, and $\Delta\varepsilon_i=\varepsilon_i-\varepsilon_\mathrm{g}$ is the energy difference between the excited state $i$ and the ground state. 

It was previously shown that BLG quantum dots can be tuned such that a single orbital is populated at a time~\cite{garreis2021ShellFilling,moller2023UnderstandingFourfold}.
Here, we take advantage of this tunability to determine the number of underlying spin and valley micro-states, by measuring the tunnel rates of the first charge carrier into and out of a single-carrier BLG quantum dot using a charge detector.
Resolving the energy dependence of the rates then gives access to the spectrum and its magnetic field dependence.

\section{Results}
\subsection{Device}
The device is based on a stack of two-dimensional materials as illustrated on figure~\ref{fig:1}a, which was assembled using the standard dry transfer technique~\cite{wang2013OneDimensionalElectrical, yankowitz2019VanWaals, purdie2018CleaningInterfaces}. 
It consists of a $\SI{35}{nm}$ thick top hBN layer, the Bernal BLG sheet, a $\SI{28}{nm}$ thick bottom hBN, and a graphite backgate layer. 
The BLG was subsequently contacted with 1D edge contacts~\cite{wang2013OneDimensionalElectrical}. % (using poly-bisphenol(A)-carbonate)
Three metallic top gates ($\SI{3}{nm}$ Cr, $\SI{20}{nm}$ Au) were evaporated on top of the stack to define two conducting channels. 
Most notable is the central split gate (labeled sC in figure~\ref{fig:1}b), which has a nominal width of $\SI{140}{nm}$ at the narrowest point. 
An additional set of top gates ($\SI{3}{nm}$ Cr, $\SI{20}{nm}$ Au) was also deposited, on top of a $\SI{26}{nm}$ thick insulating layer of aluminum oxide deposited by atomic layer deposition. 
For this experiment, the backgate is polarized at $\SI{-7.25}{V}$ and the split gates (labeled sT, sC, and sB in figure~\ref{fig:1}b) at around $\SI{8.4}{V}$, yielding a displacement field of $\SI{0.9}{V/nm}$ (assuming a relative dielectric constant of 3.5 for the hBN).
At such displacement fields, the bandgap in BLG was previously measured to be $\sim \SI{100}{meV}$~\cite{zhang2009DirectObservation,icking2022TransportSpectroscopy}.
%(which, notably, is comparable to the gap achieved in InAs nanowires~\cite{barker2019IndividuallyAddressable}, where intrinsic tunnel rates as low as a few tens of hertz can be reached~\cite{barker2022ExperimentalVerification}).  
This arrangement of patterned gates enables us to form two capacitively coupled quantum dots, one of which acts as a charge detector for the other~\cite{field1993MeasurementsCoulomba,sprinzak2002ChargeDistribution,kurzmann2019ChargeDetection}. 

\subsection{Time-resolved measurement of the excited states}\label{sec2}

\begin{figure*}
    \centering
    \includegraphics[width=\textwidth]{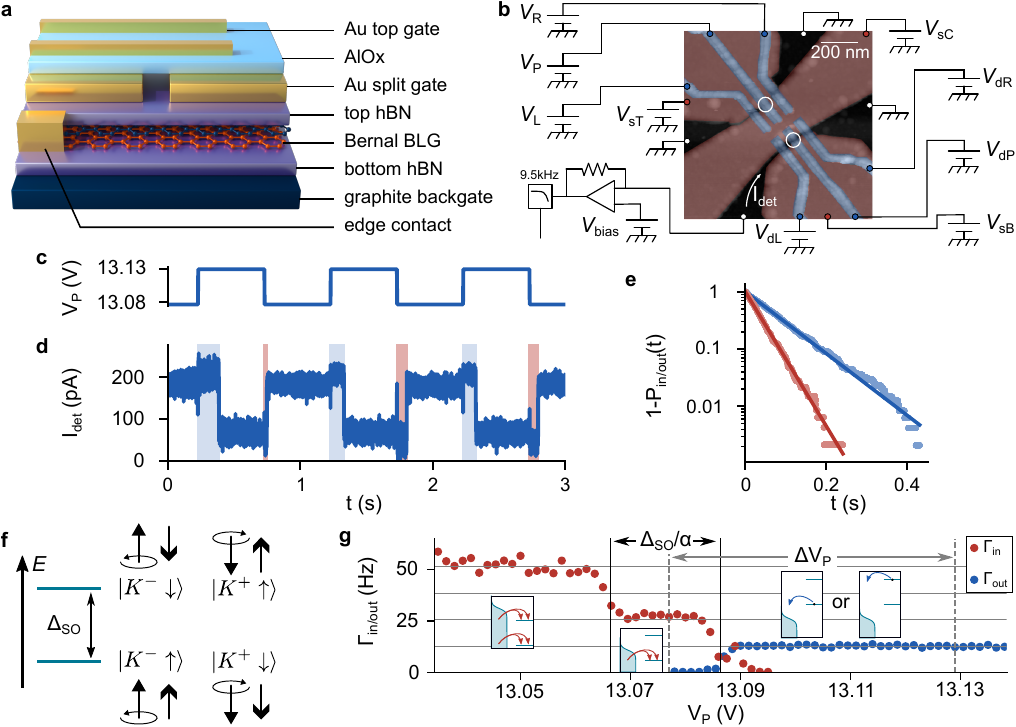} 
    \caption{Time-resolved charge detection and tunneling spectroscopy in a bilayer graphene quantum dot. \textbf{a} 3D schematic of the stack \textbf{b} False-colored atomic force microscopy image of the circuit and schematic of the measurement apparatus: a global backgate p-dopes the BLG where metallic top gates (in red) are patterned and polarized to form two conductive channels (dark areas). 
    A second layer of top gates (in blue) enables the formation of a quantum dot in each channel, their location is indicated by white circles. The top left one is the system-dot, while the bottom right one is used as a detector.
    \textbf{c} The plunger gate voltage $V_\mathrm{P}(t)$ drive and \textbf{d} the simultaneously acquired detector current $I_\mathrm{det}(t)$ [marked in (b)]. The red (blue) shaded areas correspond to individual tunneling-in(-out) times.
    \textbf{e} Plot of the probability of a single hole to have already tunneled into (out) of the dot as a function of the waiting time, as red (blue) data points. The continuous lines represent the exponential probability expected from the average waiting times.
    \textbf{f} Schematic of the spin-valley locked states constituting the Kramers pairs, separated in energy by the spin-orbit coupling. The sketches depict the valley and spin magnetic momenta.
    \textbf{g} Average tunneling-in(-out) rates in red (blue) as a function of the plunger gate voltage $V_\mathrm{P}$. The two vertical black lines delimit the extent of the $\Gamma_\mathrm{in}=2\Gamma$ plateau. The two vertical dashed gray lines indicate the two values of $V_\mathrm{P}$ at which the data of (c--e) were acquired.
    The insets are schematics representing the different tunneling possibilities in and out of a dot with two levels that are each two-fold degenerate.
    }\label{fig:1} 
\end{figure*}

The tunnel rates were measured by driving the dot's occupancy between the two charge states: 0 and 1 hole. 
The nearby charge detector (represented by the bottom right white circle in figure~\ref{fig:1}b) enables us to acquire statistics on the random time taken by the charge carrier to tunnel into (out of) the dot (top left white circle in figure~\ref{fig:1}b).
The bias across the detector-dot $V_\mathrm{bias}=\SI{14}{\micro V}$ was kept small on purpose to avoid photo-assisted tunneling~\cite{leturcq2008FrequencyselectiveSinglephoton} and heating while the system-dot remained unbiased and its barriers were set to minimize the tunnel rates. 
A periodic square drive with amplitude $\Delta V_\mathrm{P}=\SI{52}{mV}$ and frequency $\SI{1}{Hz}$, was applied to the system-dot's plunger gate (labeled with $V_P$ in figure~\ref{fig:1}b), such that the dot's most stable state alternated between single hole occupancy and empty.
The detector current was channeled through an analog low-pass filter with a $\SI{9.5}{kHz}$ bandwidth to avoid any aliasing effect at our recording sampling frequency of $\SI{20}{kHz}$. 
The detector current was then digitally filtered with a notch filter centered at $\SI{698}{Hz}$ to reduce the triboelectric noise originating from vibrations of the dilution refrigerator's pulse tube.
In addition, a moving median filter spanning 15 points (corresponding to a $\SI{0.75}{ms}$ time window) was utilized to enhance our signal-to-noise ratio, while keeping a sharp response of the detector current.
An example of such a processed detector current trace is plotted in figure~\ref{fig:1}d, concurrently with the plunger gate drive in figure~\ref{fig:1}c. 
Figure~\ref{fig:1}d illustrates that the detector current jumps between two levels, centered around $\SI{60}{pA}$ and $\SI{180}{pA}$, corresponding to the first hole being out of and in the dot, respectively.
By comparing the gate drive (figure~\ref{fig:1}c) and the detector's response (figure~\ref{fig:1}d), it is clear that the tunnel events are not synchronized with the gate drive but rather occur with some delay. 
The delays for each individual event were measured and represented by the blue (red) shaded areas in figure~\ref{fig:1}d. 
For each configuration, we acquired 82 times $\SI{6}{s}$ long traces, adding up to $\sim$450 events, enabling us to obtain a reliable average waiting time for a charge carrier to tunnel into and out of the dot.
It is possible to deduce the tunnel rates from the inverse of these average waiting times (see Supplementary Information for how we account for the finite time measurement window on the average rates).
Importantly, as the sequential tunneling of a charge to or from a dot is a Poisson process, the distributions of waiting times are expected to be exponential. 
More specifically, the probability for the charge carrier to have tunneled in (out) after time $t$ should follow $P_\mathrm{in(out)}(t) = 1-\exp{ [\Gamma_\mathrm{in(out)} t]}$.
Therefore, we collected the waiting times for each type of event, tunneling in and out. 
In figure~\ref{fig:1}e, we plot the quantity $1-n_\mathrm{in(out)}(t)/N_\mathrm{in(out)}$ which corresponds to $1-P_\mathrm{in(out)}(t)$, where $n_\mathrm{in(out)}(t)$ is the number of in (out) tunneling events that occurred after time $t$, and $N_\mathrm{in(out)}$ is the total number of tunneling in (out) events. 
The solid lines in figure~\ref{fig:1}e depict the exponential laws obtained from the inverse average waiting times, showing excellent agreement with the expected exponential decay of $1-P_\mathrm{in(out)}$. 
%We thereby validate that the tunneling-in (-out) rates $ \Gamma_\mathrm{in(out)} $ can be reliably extracted from the inverse average waiting times for the charge to tunnel in (out of) the dot. 

We now turn to the dependence of the measured rates on the plunger gate voltage $V_\mathrm{P}$. 
We applied the previously described scheme for measuring the tunnel rates in addition to varying the plunger gate voltage, keeping the amplitude $\Delta V_\mathrm{P}$ of the square drive constant. 
The obtained average rates are plotted in figure~\ref{fig:1}g, where it can be seen that the tunneling out rate (in blue) is constant far away from the transition as expected from equation~\eqref{eq:Gamma_ou}.
This constitutes a calibration of the intrinsic tunnel rate $\Gamma=\SI{12.9}{Hz}$. 
By contrast, there is a step in $\Gamma_\mathrm{in}$, separating two plateaus at $2.12\Gamma=\SI{27.4}{Hz}$ and $4.02\Gamma=\SI{51.7}{Hz}$, respectively consistent with two and four accessible micro-states within the corresponding energy windows. 
The extent of the plateau $\Gamma_\mathrm{in}\approx 2\Gamma$ corresponds to the  spin--orbit splitting $\Delta_\mathrm{SO}$, which lifts the degeneracy between the two spin-valley locked states of each of the Kramers pairs~\cite{banszerus2021SpinvalleyCoupling,kurzmann2021KondoEffect}. 
By multiplying the extent of this plateau by the independently characterized plunger gate lever arm $\alpha=0.003$, we obtain the value of the spin--orbit splitting $\Delta_\mathrm{SO}\approx 61\pm4~\mu$eV. 
Our uncertainty results from thermal broadening (see Methods).  
This value is compatible with previously reported values ranging between 40$~\mu$eV and 80$~\mu$eV~\cite{banszerus2020ObservationSpinOrbit,banszerus2021SpinvalleyCoupling,kurzmann2021KondoEffect}. 
In contrast to these standard bias spectroscopy techniques, that only gives information on the excited states energy, our method additionally gives a direct measurement of each of the states' degeneracy. 
%However, all of these were extrapolated from high magnetic field data, while here we directly measured it at zero magnetic field. 

\subsection{Magnetic field dependence}\label{sec3}
\begin{figure*}
    \centering
    \includegraphics[width=\textwidth]{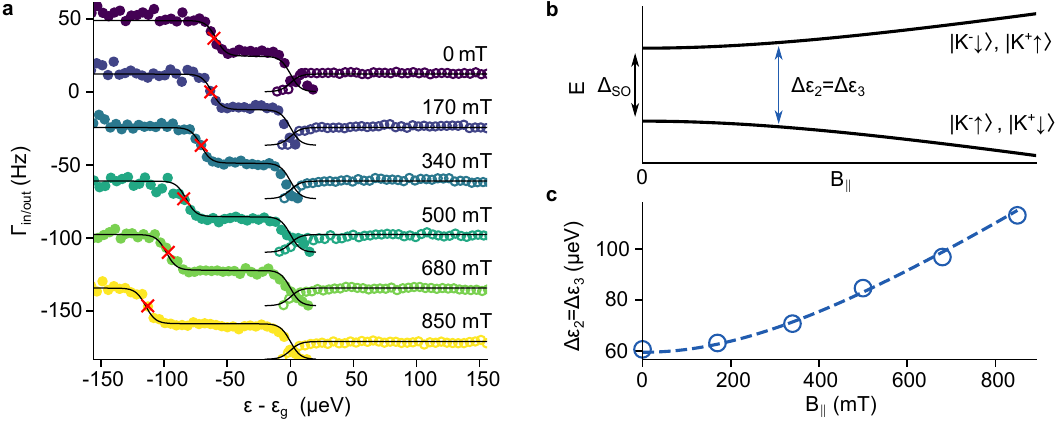}
    \caption{Tunneling rate spectroscopy in presence of an in-plane magnetic field. \textbf{a} Measured tunneling-in(-out) rates for different parallel magnetic fields. 
    Each set of curves is plotted in a distinct color and with an incremental offset of $3\Gamma$ for clarity.
    The tunneling-in(-out) rates correspond to the full (open) symbols.
    \textbf{b} Qualitative picture of the expected single-carrier excited spectrum upon applying an in-plane magnetic field.
    \textbf{c} Parallel magnetic field dependence of the single-carrier spectrum. 
    The data points are directly obtained from the red crosses of panel (a) where $\Delta\varepsilon_2=\Delta\varepsilon_3$, with uncertainty in energy given by the diameter of the circles. 
    The dashed blue line is the result of a least-square optimization carried out on the data points, taking equation~\eqref{eq:exc2} as a model and with $\Delta_\mathrm{SO}$ as a unique fit parameter. 
        \label{fig:2}}
\end{figure*}

In the following, we focus on the magnetic field dependence of the excited state spectrum. 
The predicted energy separations between the ground state $|K^-\uparrow\rangle$ and each of the three excited states $|K^+\downarrow\rangle$, $|K^-\downarrow\rangle$, and $|K^+\uparrow\rangle$ are given by \cite{knothe2022TunnelingTheory}
\begin{align}
    \Delta\varepsilon_1 &= (g_\mathrm{v}+g_\mathrm{s})\mu_\mathrm{B}B_\perp ,\label{eq:exc1} \\
    \Delta\varepsilon_2 &= \sqrt{[\Delta_\mathrm{SO} + g_\mathrm{s}\mu_\mathrm{B}B_\perp]^2 + (g_\mathrm{s}\mu_\mathrm{B}B_\parallel)^2},\label{eq:exc2} \\
    \Delta\varepsilon_3 &= \sqrt{[\Delta_\mathrm{SO} + g_\mathrm{v}\mu_\mathrm{B}B_\perp]^2 + (g_\mathrm{s}\mu_\mathrm{B}B_\parallel)^2},\label{eq:exc3} 
\end{align} 
where $g_\mathrm{v}$ is the valley Landé factor, $g_\mathrm{s}=2$ the spin Landé factor, $\mu_B$ the Bohr magneton, and $B_{\perp(\parallel)}$ the magnitude of the perpendicular (parallel) magnetic field.

To further establish the obtained value of $\Delta_\mathrm{SO}$ at zero magnetic fields, we study the parallel magnetic field dependence of the excited state spectrum, keeping $B_\perp=0$. 
The full (open) circles in figure~\ref{fig:2}a are the measured $ \Gamma_\mathrm{in(out)}$ for several values of parallel magnetic fields, where the extent of the plateau $\Gamma_\mathrm{in}=2\Gamma$ broadens as the magnetic field increases. 
The value of the intrinsic tunnel rate $\Gamma\approx \SI{12.2}{Hz}$ was first characterized by fitting each measured $\Gamma_\mathrm{in(out)}(\varepsilon)$ to equation~\eqref{eq:Gamma_ou}, and averaging the obtained values for $\Gamma$.
Then, the position of the ground state $\varepsilon_\mathrm{g}$ was established at each magnetic field (see Supplementary information for details on the ground state determination). 
We then identified the point $\Delta\varepsilon_2=\Delta\varepsilon_3$ where $\Gamma_\mathrm{out}=3\Gamma$ (see Supplementary information) for each magnetic field value and emphasized these points on figure~\ref{fig:2}a with red crosses.  
The energy differences $\Delta\varepsilon_2=\Delta\varepsilon_3$ are plotted as the blue circles in figure~\ref{fig:2}b. 
The dashed line is a fit to the expected splitting as given by \cref{eq:exc2,eq:exc3}, which simplify to $\Delta\epsilon_2 = \Delta\epsilon_3 = \sqrt{\Delta_\mathrm{SO}^2 + (g_s\mu_B B_\parallel)^2} $, in absence of a perpendicular magnetic field. 
Here, the only free parameter is $\Delta_\mathrm{SO}\approx\SI{59}{\micro eV}$, compatible with the value measured at zero magnetic field. 
As a confirmation of the previously determined values of $\Gamma$, $\varepsilon_\mathrm{g}$, and $\Delta_\mathrm{SO}$, we plot the black continuous lines in figure~\ref{fig:2}a that follow \cref{eq:Gamma_ou,eq:Gamma_in}.
The effective electronic temperature $T\approx \SI{46}{mK}$ entering in the Fermi-Dirac distribution is separately characterized from the average current values, depends on the actual electronic temperature and the dot's energy level fluctuation (see Supplementary Information) and determines the energy resolution of this method. 
These black lines are not fits to the individual sets of points, but are directly the theoretical curves of \cref{eq:Gamma_ou,eq:Gamma_in}, with all parameters independently determined.  

\begin{figure*}
    \centering
    \includegraphics[width=\textwidth]{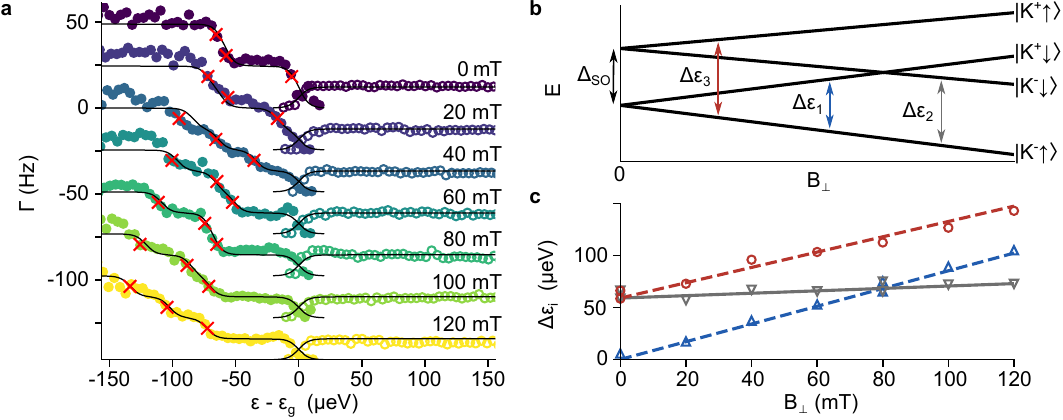}
    \caption{Tunneling rate spectroscopy in presence of an out-of-plane magnetic field. \textbf{a} Measured tunneling-in(-out) rates are plotted in full (open) symbols, for different magnetic field values each plotted in a distinct color and with an incremental offset of $2\Gamma$ for clarity.
    \textbf{b} Qualitative picture of the expected single-carrier excited spectrum upon applying a perpendicular magnetic field.
    \textbf{c} Perpendicular magnetic field dependence of the single-carrier spectrum. The data points are directly obtained from the red crosses of panel (a). They are subsequently separated into three categories, each corresponding to an excited state: the blue upward triangles correspond to excited state 1, the gray downward triangles to excited state 2, and the red circles to excited state 3. 
    The uncertainty in energy is given by the size of the symbols.
    The continuous gray line corresponds to equation~\eqref{eq:exc2}, with $B_{\parallel}=\SI{0}{T}$, $\Delta_\mathrm{SO}=\SI{59}{\micro eV}$ and $g_\mathrm{s}=2$.
    The dashed blue and red lines correspond to the result of a least square optimization on $g_\mathrm{v}$ simultaneously performed on \eqref{eq:exc1} and \eqref{eq:exc3}, to the data set with matching colors. 
        }
    \label{fig:3}
\end{figure*}

Finally, we look at the perpendicular magnetic field dependence of the spectrum, for which it is expected that all degeneracies are lifted.
The tunneling rates $\Gamma_\mathrm{in(out)}$ were measured in steps of $\SI{20}{mT}$ up to $\SI{120}{mT}$, and are shown as full (open) circles in figure~\ref{fig:3}a. 
We used the same procedure as before to determine $\Delta\varepsilon_i$, by first characterizing $\varepsilon_\mathrm{g}$ and then identifying the points where $\Gamma_\mathrm{in}=\{1.5,2.5,3.5\}\Gamma$, indicated by the red crosses in figure~\ref{fig:3}a. 
The resulting values of $\Delta\varepsilon_i$ are plotted as a function of the perpendicular magnetic field in figure~\ref{fig:3}b. 
The points are then separated into three categories, differentiated by shapes and colors in figure~\ref{fig:3}b. 
Each set of points follows a linear dependence on $B_\perp$. 
The points located at the crossings are attributed to both possible sets.
We subsequently plot the theoretical line for equation~\eqref{eq:exc2} that requires no fitting parameter, as a continuous gray line, using the previously determined $\Delta_\mathrm{SO}\approx \SI{59}{\micro eV}$, showing excellent agreement with the gray downward triangles.  
The two remaining sets correspond to the first and third excited states (\cref{eq:exc1,eq:exc3}), and require only one additional parameter $g_\mathrm{v}$, which is adjusted simultaneously for both sets. 
The red and blue dashed lines are the results of the parameter optimization for equations~\eqref{eq:exc1} and \eqref{eq:exc3}, giving $g_\mathrm{v}\approx 12.7$. 

This value of the valley Landé factor is lower than previously reported values, which range from 18 to 90~\cite{eich2018SpinValley,kurzmann2019ExcitedStates,tong2021TunableValley,moller2021ProbingTwoElectron,moller2023UnderstandingFourfold}. 
This discrepancy could be explained by the $g_\mathrm{v}$ dependence on the displacement field strength~\cite{tong2021TunableValley,moller2021ProbingTwoElectron,moller2023UnderstandingFourfold} as we here operated at a large value of $D\approx\SI{0.9}{V/nm}$, or by an unusually elongated shape of the quantum dot~\cite{tong2021TunableValley}. 
We then used this value of $g_\mathrm{v}$ as well as the previously determined parameters in~\cref{eq:Gamma_ou,eq:Gamma_in} to plot the continuous black lines on figure~\ref{fig:3}a. 
The slight deviations occurring at higher energies could be due to an energy-dependence of the barrier, which in turn would be magnetic-field dependent, as observed in~\cite{hofmann2016MeasuringDegeneracy}. 
Noticeably, there is no valley mixing of the states up to our experimental resolution. 
This attests of the absence of crystalline defects in the bilayer graphene, over the spatial extent of the quantum dot's wavefunction.
We thus establish a new upper bound on the states valley-mixing of $\SI{4}{\micro eV}$ (limited by the electronic temperature in our device), five times lower than the previously reported value~\cite{banszerus2021SpinvalleyCoupling}.

\section{Conclusion}\label{sec4}
In summary, we measured the spectrum of a single-carrier bilayer graphene quantum dot with an energy resolution of $\SI{4}{\micro eV}$.
We used a time-resolved charge detection technique that provides direct access to the number of states in the system, and therefore to their degeneracy.
This enabled us to resolve all four (ground and excited) states of a single-carrier quantum dot and their behavior as a function of both perpendicular and parallel magnetic fields.
On the technological side, we demonstrated that the tunneling rates into and out of a bilayer graphene quantum dot could be made as low as $\sim\SI{12}{Hz}$ even in the absence of a magnetic field.
Finally, our measurement scheme opens the door to measuring the lifetimes of the various excited states of different spin and/or valley nature in bilayer graphene quantum dots, close to zero magnetic fields~\cite{denisov2024UltralongRelaxation}.

\appendix

\section{Methods}\label{sec_Methods}
\subsection{Effective electronic temperature}
The electronic temperature is obtained from the value of the detector current averaged over the last $\SI{0.3(0.05)}{s}$ of each pulse, for a gate frequency of $f_\mathrm{VP}=\SI{1(2)}{Hz}$, to ensure the dot is as close as possible to equilibrium.
%In a first step, we determine the slope of the detector current coming from the 
Each average current curve is then fitted to a slanted Fermi distribution
\begin{equation}\label{eq:fermi}
     \frac{ a I_e}{ 1+\exp \left[ \alpha e(V_\mathrm{P}-V_0)/k_B T_\mathrm{e}\right] } + I_\mathrm{off} 
\end{equation}
with $a$ the slope (corresponding to the cross-talk of $V_\mathrm{P}$ on the detector), $I_e$ the step height, $T_\mathrm{e}$ the electronic temperature, $V_0$ the center of the transition and $I_\mathrm{off}$ an offset contribution, which are all fitting parameters.

\begin{figure}
    \centering
    \includegraphics{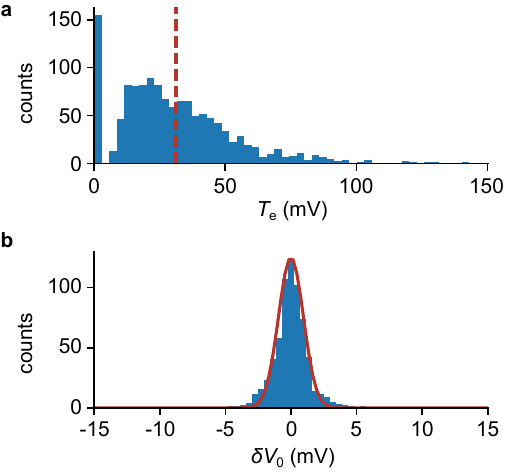}
    \caption{Histograms of the results of the fits of the average detector current to equation \eqref{eq:fermi}, for all our gathered data, for two parameters. 
    (a) Histogram of the $T_\mathrm{e}$ parameter.
    The dashed red vertical line indicates the median value $T_\mathrm{e} \approx \SI{31.2}{mK}$ (ignoring the first bin close to $T_\mathrm{e}=0$).
    (b) Histogram of the  $\delta V_0 = V_0 - \langle V_0(B) \rangle$, where $V_0$ is the fit parameter and $\langle V_0(B) \rangle$ is the average for all plunger gate sweeps at a given magnetic field.
    The continuous red line is a Gaussian distribution with a standard deviation of $\sigma_{V0} = \SI{0.95}{mV}$. }
    \label{fig:SI_Temperature}
\end{figure}

A histogram of all obtained $T_\mathrm{e}$ for all our gathered data sets used in the main text are plotted in figure \ref{fig:SI_Temperature}a.
The median value is $\SI{27.4}{mK}$.
We can clearly see that the bin close to $T_\mathrm{e} = 0$ is the most prominent. 
We understand this as resulting from sudden jumps of the transition that have an amplitude on the order of $k_B T_\mathrm{e}$, over the course of our plunger gate sweep. 
As these artificially reduce the median, we exclude the first bin to obtain a more reliable value of the median of $T_\mathrm{e} \approx \SI{31.2}{mK}$, which we then use.

To take into account the fast jumps as well as the slow drifts of the transition in the effective temperature, we also plot a histogram of the transition centers variation $\delta V_0 = V_0 - \langle V_0(B) \rangle$, with $V_0$ the transition center of an individual sweep and $\langle V_0(B) \rangle$ the average of the transition centers at each magnetic field, in figure \ref{fig:SI_Temperature}b. 
We can see that the transition fluctuates according to a Gaussian probability distribution with a standard deviation of $\sigma_{V0} = \SI{0.95}{mV}$ (red line in \ref{fig:SI_Temperature}b).
These fluctuations can be attributed to charge noise in the vicinity of the dot which affects the dot's electrochemical potential.

To estimate an effective temperature, we consider the slow charge noise to be independent of the temperature extracted from the Fermi fits and we obtain $T \approx \sqrt{T_\mathrm{e}^2 + (e \alpha \sigma_{V0}/k_B)^2 } \approx \SI{46}{mK}$, which we use in the main text.

\subsection{Digitization procedure}
To quantify the lifetimes, we first digitize the measured detector current on two levels.
In the first step, we fit the histograms of the current values with two Gaussian distributions to obtain the average current value of each level $\langle I_\mathrm{up} \rangle$ and $\langle I_\mathrm{dn} \rangle$.
Second, we define two thresholds $I_\mathrm{th-up}$ and $I_\mathrm{th-dn}$ such that 
\begin{align}
    I_\mathrm{th-up} &= \langle I_\mathrm{up} \rangle - \gamma (\langle I_\mathrm{up} \rangle-\langle I_\mathrm{dn} \rangle)/2 \\
    I_\mathrm{th-dn} &= \langle I_\mathrm{dn} \rangle + \gamma (\langle I_\mathrm{up} \rangle-\langle I_\mathrm{dn} \rangle)/2
\end{align} 
with $\gamma$ a parameter that we have to fix. 
We found that the digitization was almost identical whenever $0\leq \gamma \leq 0.4$ for all of our data. 
We therefore used the value $\gamma=0.2$ for digitizing all of the traces acquired for this work. 
Third, we apply an algorithm that scans each point sequentially and attributes each point to a category as follows.
Any point above $I_\mathrm{th-up}$ is digitized as a 1, any point below $I_\mathrm{th-dn}$ is digitized as a 0, and any point located in between the two thresholds is attributed the same value as the previous point. 
(For the edge case where the first few points of a trace lie in between the two thresholds, those points are ignored).
This double threshold method enables us to reliably digitize our traces that have a signal-to-noise ratio of $\mathrm{SNR}\approx (\langle I_\mathrm{up} \rangle - \langle I_\mathrm{dn} \rangle)/[(\sigma_\mathrm{up}+\sigma_\mathrm{dn})/2] \approx 5.3$ (for the data displayed in figure 1c of the main text, which is representative of all our data).
Note that this SNR could be made larger by using a larger detector bias, but is voluntarily kept relatively small to avoid photo-assisted tunneling in our system dot. 

\subsection{Delay on the gate pulse}
The plunger gate voltage $V_\mathrm{P}$ is controlled by a voltage source that has a characteristic time of $\sim 10^{-6}$~s, and through a line that is low-pass filtered with a simple RC circuit, which has a characteristic time of $10^{-5}$~s. 
The gate response time is therefore negligible in comparison to our measured lifetimes.
It is in fact possible to observe this directly in the data: in figure 1c of the main text, there are very sharp peaks in the detector current, that are synchronized with the gate pulses. 
Those are the signature of the high-pass filter behaviour of the system-dot's plunger gate on the detector current, and it can be observed that this time scale is much smaller than that of the lifetimes. 

\subsection{Determination of the lifetimes}
We here explain in more detail how we extract each individual lifetime.
Each period of the plunger gate drive is constituted of two constant--voltage half-periods as shown in figure 1c of the main text.
Each of these half-periods is cut in 4 different time windows: 
\begin{itemize}
    \item the begin edge time window (the first $\SI{1.1}{ms}$) ,
    \item the checking time window (the subsequent $\SI{1}{ms}$)   
    \item the measurement time window (remaining of the time up to the last window) 
    \item the end edge time window (the last $\SI{1.1}{ms}$) 
\end{itemize}
The two edge time windows, at the beginning and at the end of every half period, are cut out of the analysis to avoid false counting due to the cross-talk of the plunger gate on the detector current (see previous Methods section).
The checking time window consists in monitoring the state of the system-dot and checking that it is in its expected initial state and remains so for this small amount of time. 
If it is not, or if it switches during this time, the events occurring during the whole half-period are not considered in our statistics.
Finally, the measurement time window starts at a time defined as $t_0$: the beginning of our measurement of the lifetimes. 
The state of the dot is monitored for this whole time and we record the time $t_\mathrm{switch}$ when the state of the system-dot switches for the first time during this time window.
The lifetime is then defined as $\tau = t_\mathrm{switch}-t_0$.
Only the first switching event is considered, all other subsequent events occurring in the half-period are ignored.
In this way, we circumvent the difficulty that arises in the analysis of telegraph noise, for which the state of the system-dot is not well known at any time.
On the contrary, with the method we use, the system-dot state is fully known at time $t_0$, which then gives us access to the actual dot occupation probability over time.

\section{Data availability}
The data generated in this study have been deposited in the ETH Research Collection database under accession code https://doi.org/10.3929/ethz-b-000686402

\section{Acknowledgements}
H.D. would like to thank Olivier Maillet, Everton Arrighi, Rebeca Ribeiro-Palau, and Artem Denisov for interesting discussions and suggestions. We also thank Lin Wang for pointing a mistake in the preprint. We acknowledge financial support by the European Graphene Flagship Core3 Project, H2020 European Research Council (ERC) Synergy Grant under Grant Agreement 951541, the European Innovation Council under grant agreement number 101046231/FantastiCOF, NCCR QSIT (Swiss National Science Foundation, grant number 51NF40-185902).
K.W. and T.T. acknowledge support from the JSPS KAKENHI (Grant Numbers 21H05233 and 23H02052) and World Premier International Research Center Initiative (WPI), MEXT, Japan.

\section{Authors contributions Statement}
T.T., and K.W. grew the hBN.
H.D. fabricated the device with inputs from M.M. and M.R..
A.O. and H.D. characterized and tuned the dots. 
S.C. and H.D. acquired and analyzed the data with inputs from T.I. and K.E..  
H.D. wrote the manuscript with inputs from S.C., A.O., M.M., M.R., C.A., C.T., R.G., J.G., W.H., L.G., T.I., and K.E.

\end{document}